# Intrinsic magnetic properties of the layered antiferromagnet CrSBr

Fangchao Long[1,2], Kseniia Mosina[3], René Hübner[1], Zdenek Sofer[3], Julian Klein[4], Slawomir Prucnal[1], Manfred Helm[1,2], Florian Dirnberger[5,*], Shengqiang Zhou[1,*]

[1]Helmholtz-Zentrum Dresden-Rossendorf, Institute of Ion Beam Physics and Materials Research, Bautzner Landstrasse 400, 01328 Dresden, Germany
[2]TU Dresden, 01062 Dresden, Germany
[3]Department of Inorganic Chemistry, University of Chemistry and Technology Prague, Technická 5, 166 28 Prague 6, Czech Republic
[4]Department of Materials Science and Engineering, Massachusetts Institute of Technology, Cambridge, Massachusetts 02139, USA
[5]Institute of Applied Physics and Würzburg-Dresden Cluster of Excellence ct.qmat, TU Dresden, 01062 Dresden, Germany

[*]Corresponding authors: florian.dirnberger@tu-dresden.de, s.zhou@hzdr.de

## Abstract

Van der Waals magnetic materials are an ideal platform to study low-dimensional magnetism. Opposed to other members of this family, the magnetic semiconductor CrSBr is highly resistant to degradation in air, which, besides its exceptional optical, electronic, and magnetic properties, is the reason the compound is receiving considerable attention at the moment. For many years, its magnetic phase diagram seemed to be well-understood. Recently, however, several groups observed a magnetic transition in magnetometry measurements at temperatures of around 40 K that is not expected from theoretical considerations, causing a debate about the intrinsic magnetic properties of the material. In this letter, we report the absence of this particular transition in magnetization measurements conducted on high-quality CrSBr crystals, attesting to the extrinsic nature of the low-temperature magnetic phase observed in other works. Our magnetometry results obtained from large bulk crystals are in very good agreement with the magnetic phase diagram of CrSBr previously predicted by the mean-field theory; A-type antiferromagnetic order is the only phase observed below the Néel temperature at $T_N = 131$ K. Moreover, numerical fits based on the Curie-Weiss law confirm that strong ferromagnetic correlations are present within individual layers even at temperatures much larger than $T_N$.

## Main

Van der Waals (vdW) magnetic materials are an interesting platform for fundamental research and potential technological applications in magneto-optics and spin-based electronics.[1-4] Since magnetic signatures were first discovered in monolayers of $CrI_3$[5] and bilayers of $Cr_2Ge_2Te_6$[6], issues related to the degradation of these particular materials in air have been a challenge for sample fabrication. In search of alternatives, other vdW magnetic compounds less sensitive to degradation in air are being investigated at the moment.

Within this family of air-stable magnetic vdW crystals, the semiconductor CrSBr is of particular importance.[7-11] Early work reported its A-type antiferromagnetic (AFM) order in which the magnetization of individual layers with intralayer ferromagnetic (FM) coupling alternates in the stacking direction.[12] In multilayer samples, AFM coupling between the layers is observed up to the Néel temperature ($T_N = 132 \pm 1$ K) while strong FM correlations couple Cr ions within each layer at even higher temperatures. As a result, the magnetic transition temperature of a single layer of CrSBr is higher than the Néel temperature in bulk. It was observed to be around 160 K,[13-15] which is also substantially larger than the magnetic transition temperatures found in $CrI_3$[5] and $Cr_2Ge_2Te_6$[6]. Although earlier work on CrSBr reported only the AFM order below 132 K[12,16], a second magnetic phase with FM signatures was recently observed in magnetometry measurements below 40 K in crystals grown by different groups,[13,15,17-20] leading to a debate about its origin. Telford *et al.*[13,15], for example, suggest an extrinsic nature related to magnetic defects, an idea further corroborated by the work of Klein *et al.*[17] Other groups, however, concluded that this magnetic phase represents the intrinsic magnetic properties of CrSBr. Wu *et al.*[18] attribute it to incoherently coupled 1D electronic chains, while López-Paz *et al.*[19] relate the low-temperature phase transition to a slowing-down of magnetic fluctuations, intrinsically driven by the in-plane uniaxial anisotropy of CrSBr. Boix-Constant *et al.*[20] further considered this magnetic phase to represent a cooperative state in which the spins (both within and between the layers) are fully frozen along the direction of an external magnetic field. Besides that, we recently showed that the irradiation of CrSBr with high-energy ions can result in a permanent change of the magnetic order from AFM to FM.[21]

In this study, we demonstrate the absence of the second magnetic phase in magnetometry measurements of high-quality CrSBr crystals. Precluding an intrinsic origin of the 40 K-state, our results directly address questions raised in other works. Down to temperatures of 2 K, both DC and AC magnetometry measurements show no signatures other than those related to the AFM order. A numerical analysis based on Curie-Weiss fits further indicates that intralayer FM correlations are present beyond $T_N$.

High-quality CrSBr crystals were synthesized by a direct reaction from the elements. Chromium, bromine, and sulfur were mixed in stoichiometric ratio in a quartz ampoule. The material was pre-reacted in an ampoule using a crucible furnace at 700 °C for 12 hours, while the second end of the ampoule was kept below 250 °C. The heating procedure was repeated two times until the liquid bromine disappeared. The ampoule was placed in a horizontal two-zone furnace for crystal growth. First, the growth zone was heated to 900 °C, while the source zone was heated to 700 °C for 25 hours. For the growth, the thermal gradient was reversed and the source zone was heated from 900 °C to 940 °C and the growth zone from 850 °C to 800 °C over a period of 7 days. High-quality CrSBr single crystals were obtained with lengths up to 2 cm. The Raman spectrum of the as-synthesized CrSBr crystals reported herein were measured by micro-Raman spectroscopy using a linearly polarized, continuous-wave 532 nm-Nd:YAG laser for excitation. Atomic-number-contrast high-angle annular dark-field (HAADF) scanning transmission electron microscopy (STEM) imaging was performed via using a Talos F200X microscope operated at an accelerating voltage of 200 kV. The magnetization data were collected by a superconducting quantum interference device (Quantum Design, SQUID-VSM) magnetometer.

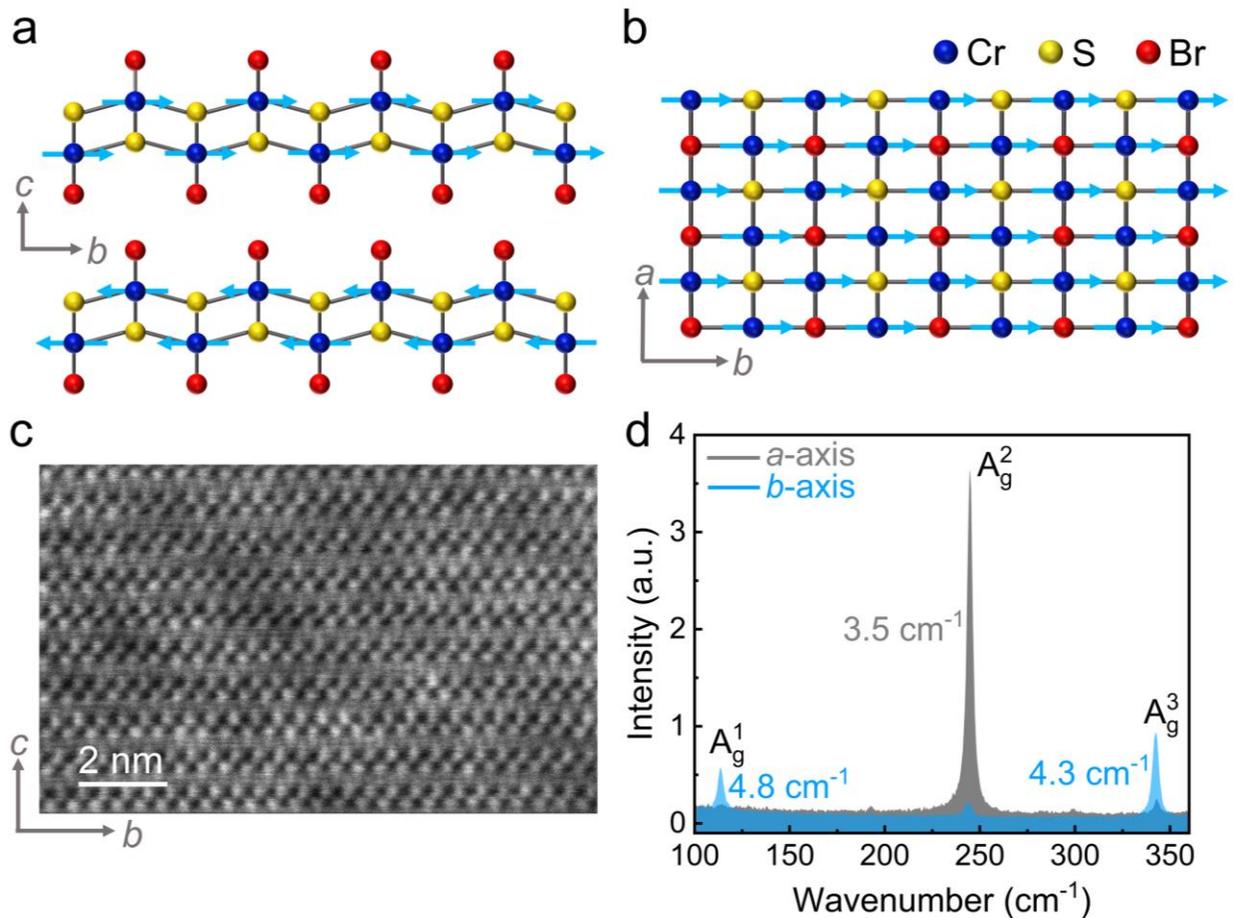

**Figure 1**. **Structural properties of CrSBr.** Schematic of the CrSBr crystal viewed along the (a) *a*-axis and (b) *c*-axis. The orientation of the Cr magnetic moments is depicted by light blue arrows. (c) Cross-sectional high-resolution HAADF-STEM image of CrSBr. d) Raman spectrum of CrSBr. The numbers are the full widths at half maximum of the corresponding peaks.

As shown in Fig. 1a, CrSBr exhibits a layered structure in which each layer consists of a buckled plane of CrS chains embedded in Br atoms. Every layer exhibits spontaneous magnetization along the *b*-axis (see Fig. 1b). The cross-sectional high-resolution scanning transmission electron microscopy (STEM) image depicted in

Fig. 1c clearly shows the layered structure of the compound. Raman measurements reveal the three characteristic Raman modes related to the out-of-plane vibration of the atoms and the strong crystal anisotropy.[22] The $A_g^1$ (113.5 cm$^{-1}$) and $A_g^3$ (342.7 cm$^{-1}$) modes observed in Fig. 1d show maximum intensity when the laser polarization is parallel to the *b*-axis, while the $A_g^2$ (244.7 cm$^{-1}$) mode is most pronounced when the polarization is along the *a*-axis. We note that the full width at half maximum (FWHM) values of our crystals (4.8 cm$^{-1}$, 3.5 cm$^{-1}$, and 4.3 cm$^{-1}$) are significantly smaller than the values reported in a recent study (16.4 cm$^{-1}$, 18.1 cm$^{-1}$, and 20.3 cm$^{-1}$ for the $A_g^1$, $A_g^2$, and $A_g^3$ modes, respectively),[15] indicating smaller structural inhomogeneity in our CrSBr crystals.

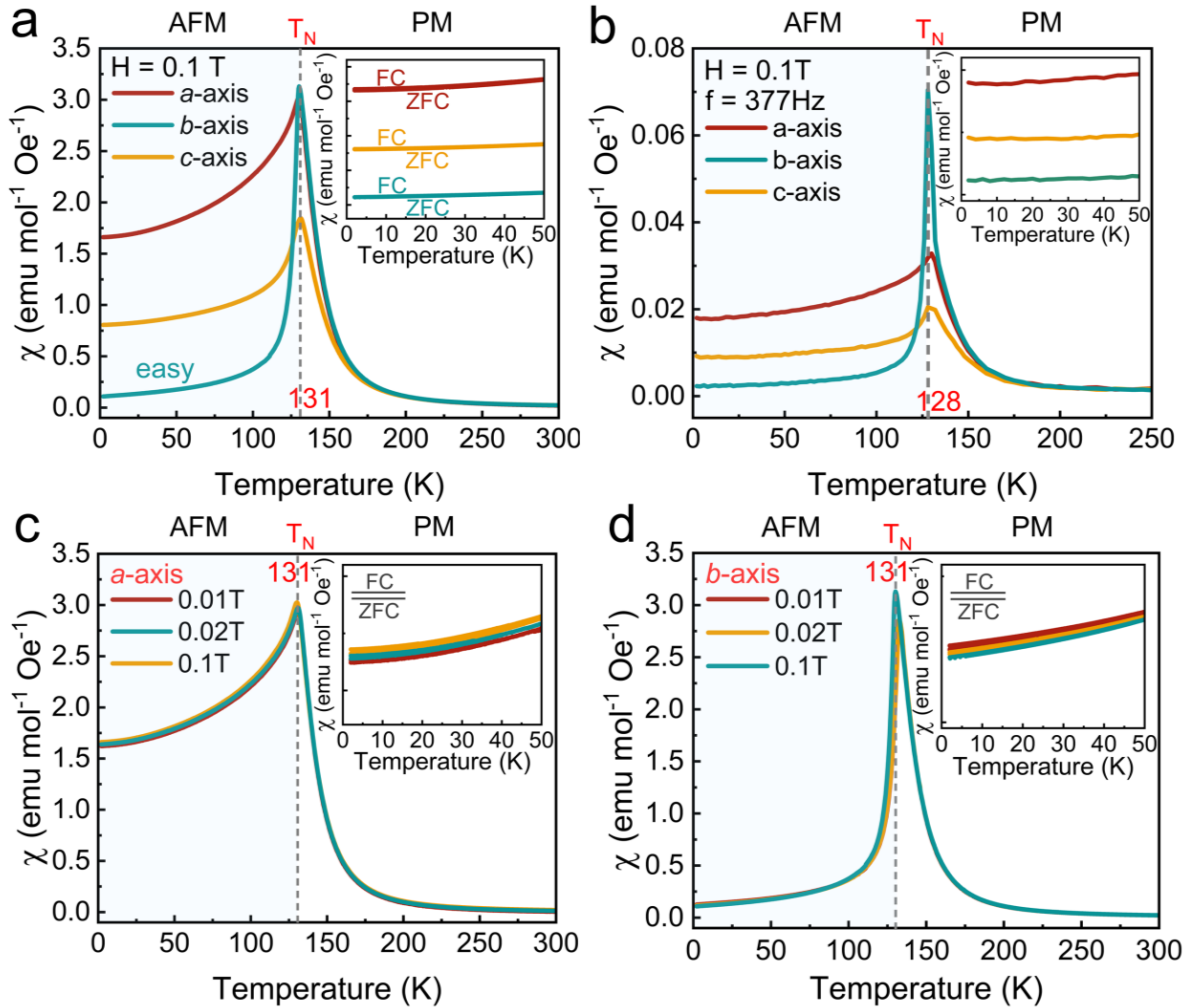

**Figure 2. Temperature-dependent magnetic properties of CrSBr.** (a) DC magnetic susceptibility and (b) AC magnetic susceptibility measurements as a function of temperature for fields applied along the *a*-axis (red), *b*-axis (green), and *c*-axis (orange). (c), (d) DC magnetic susceptibility as a function of temperature for fields applied along (c) the *a*-axis and (d) the *b*-axis with a magnitude of 0.01 T (red), 0.02 T (green), and 0.1 T (orange). The insets show the ZFC and FC curves for temperatures between 2 and 50 K. The scale of the magnetic susceptibility axis in the inset is the same as that of the main plot.

To study the magnetic properties of our CrSBr crystals, we determined the temperature- and field-dependent magnetization using DC and AC magnetometry measurements (Superconducting Quantum Interference Device, SQUID). Figure 2a shows the temperature-dependent magnetic susceptibility ($\chi$) of an as-grown CrSBr crystal for fields consecutively applied along the different principal crystallographic axes. The sharp cusp at T = 131 K marks the Néel temperature ($T_N$), where the phase transition from the paramagnetic (PM) phase to the AFM phase occurs as we cool our sample. It is important to note that we do not observe any of the experimental signatures such as kinks or discontinuities in the susceptibility data around T = 40 K that were reported by other groups. The inset of Fig. 2a compares the susceptibility curves recorded while cooling the sample without a magnetic field (labelled zero field cooling, ZFC) with those obtained in the presence of a 0.1-T-strong external field (labelled field cooling, FC). While marked differences between the ZFC and FC curves were reported in other studies, indicating the presence of FM correlations between the magnetic moments of different layers[13,15,19,20], we do not observe any noticeable difference between the two curves. The response observed in our measurements matches the expectation for pure AFM order.

To further confirm our observations, we performed additional AC susceptibility measurements using an external field of $B_{AC}$ = 1 mT alternating at a frequency of 377 Hz. This alternative measurement of the magnetic susceptibility is often more sensitive to magnetic phase transitions than DC magnetic measurements. However, as depicted in Fig. 2b, no anomalous signatures are observed below $T_N$. Overall, all curves measured along the *a*-, *b*-, and *c*-axes exhibit a shape similar to that observed in the DC measurements. The main difference is the much larger intensity of the cusp when the field is applied along the *b*-axis in the AC measurement, which we ascribe to a stronger anisotropy of the magnetization in response to low-frequency alternating fields. Magnetization measurements recorded for fields applied along the *a*- (Fig. 2c) and *b*-axis (Fig. 2d) do not show any anomalous signatures either. We note that all these measurements were repeated on two more crystals and that none of them showed any signatures of the secondary low-temperature phase observed by other groups (cf. Fig. S1).

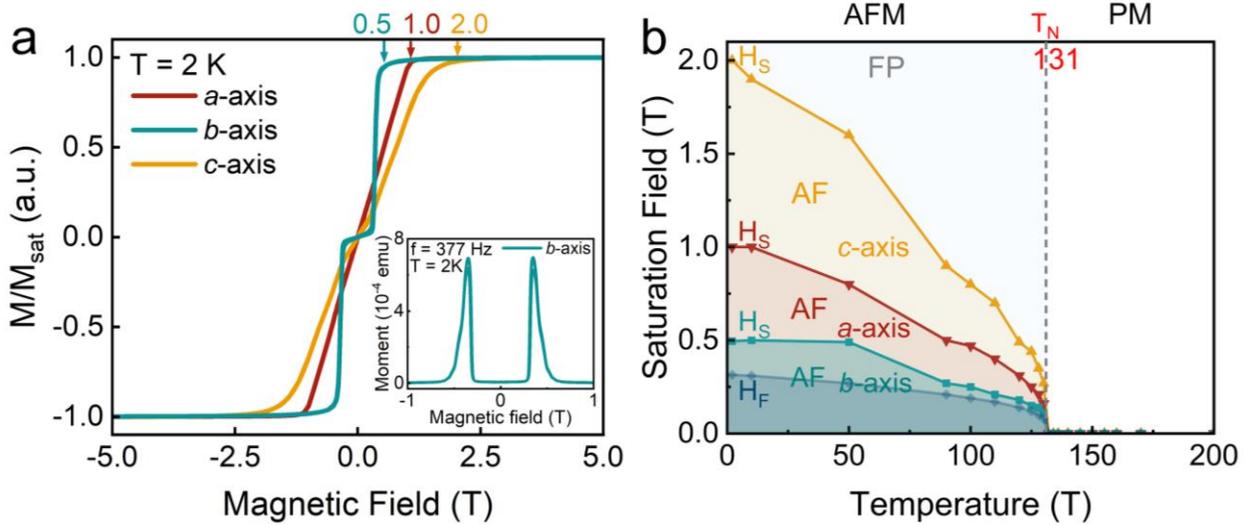

**Figure 3. Field-dependent magnetic properties of CrSBr.** (a) Magnetization at 2 K along the crystallographic *a*-, *b*-, and *c*-axes. The inset shows the AC magnetization at 2 K measured at a frequency of 377 Hz. (b) Temperature-dependent saturation field and spin-flip field.

In addition, we measured the dependence of the magnetization **M** as a function of the applied magnetic field **H** along the main crystallographic axes using SQUID magnetometry. The saturation magnetization is around 3 $\mu_B$/Cr, but slightly differs along the different axes due to the geometry effect of the SQUID magnetometer.[23] In order to show the anisotropy, the magnetization along the different axes is normalized. As illustrated in Fig. 3a, we observe the expected behavior, i.e., **M** resembles a step-like function when the field is applied along the magnetic easy axis (*b*-axis). At the saturation field, the system undergoes a first-order phase transition from AFM to FP (fully polarized). In contrast to that, **M** versus **H** curves recorded for fields applied along the *a*- or *c*-axis display a continuous increase until saturation is reached, reflecting the progressive field-induced canting of the spins along the field direction. Therefore, the *a*-, *b*-, and *c*-axis are respectively assigned to the intermediate, easy, and hard magnetization axis, which is consistent with the observation that the $\chi$ versus T curve along the *b*-axis is the sharpest. No hysteresis centered around zero field, indicative of FM correlations, is observed (see Fig. S2).

The inset in Figure 3a shows the result of AC magnetization measurements at 2 K. Only two distinct peaks are observed, both at fields of ± 0.3 T, reflecting the onset of the first-order AFM to FP phase transition. At T = 2K, the fields at which the magnetization is fully saturated are determined to be 0.5, 1.0, and 2.0 T for the *a*, *b*, and *c*-axis, respectively. The temperature-dependence of the **M** versus **H** curves (Fig. S3) shows only one transition between 120 K and 150 K at $T_N$. Figure 3b summarizes the temperature-dependent saturation fields (**H$_S$**) and the spin-flip field (**H$_F$**). As the temperature increases, **H$_S$** and **H$_F$** gradually reduce to zero up to the point where the transition from AFM to PM occurs.

In the PM phase, the temperature-dependence of $\chi$ can be described by the Curie-Weiss law:

$$\chi = \chi_0 + \frac{C}{T - \Theta_{CW}}$$

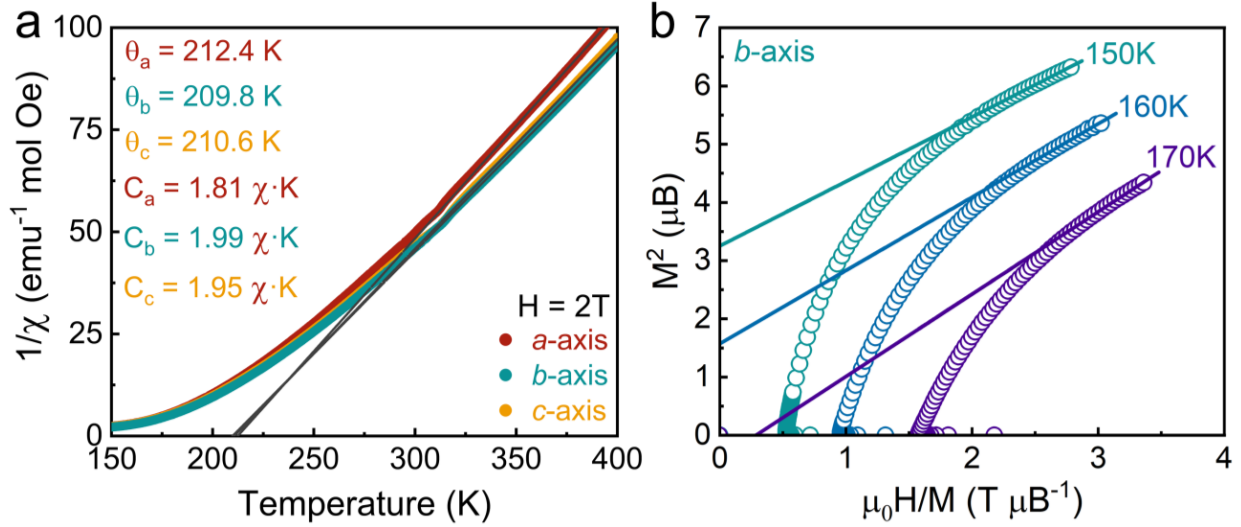

**Figure 4. Magnetic properties of CrSBr in the PM phase.** (a) Curie-Weiss fitting of the inverse magnetic susceptibility versus temperature. Linear fits to the PM regime are given by black lines. (b) Arrott plots at selected temperatures. Linear fitting at different temperatures are given by lines with corresponding color.

where $\chi_0$ is the temperature-independent susceptibility arising from core diamagnetism and the background, $C$ is the Curie constant, and $\Theta_{CW}$ is the Weiss constant. The Weiss temperature ($\Theta_{CW}$) reflects the type of magnetic interaction in a system, with $\Theta_{CW}$ being negative when there is antiferromagnetic exchange and positive when there is ferromagnetic interaction.[16] As can be seen in Fig. 4a, the extracted Weiss constants, which are 212.4 K, 209.8K, and 210.6 K for the *a*-, *b*-, and *c*- axis, respectively, are positive and much larger than $T_N$. Figure 4b illustrates the Arrott plots[24] in proximity to the paramagnetic transition, revealing a critical temperature ranging from 160 to 170 K. Such high critical temperatures, in conjunction with the large Weiss constants, indicate the strong local FM intralayer correlations that were recently also found by field- and temperature-resolved exciton spectroscopy.[10]

In conclusion, our experiments show that the low-temperature magnetic phase observed in other magnetometry studies around T = 40 K is absent in high-quality CrSBr crystals. A full characterization of the magnetic properties of these CrSBr crystals confirms the A-type AFM order below $T_N$ = 131 K predicted by the mean-field theory.[12] Beyond that, we observe strong FM intralayer correlations near the AFM-to-PM phase transition. We therefore conclude that the anomalous signatures, observed in magnetometry measurements by other groups at around 40 K, are of extrinsic origin.

See the supplementary material for the experimental detail and supporting data, including Magnetization and AC Magnetization of randomly selected CrSBr crystals; Magnetization of CrSBr at low range of the applied magnetic field; Magnetization versus magnetic field along *a*-, *b*-, and *c*-axis at different temperatures.

The authors thank Annette Kunz for TEM specimen preparation. Furthermore, the use of the HZDR ion beam center TEM facilities and the funding of TEM Talos by the German Federal Ministry of Education and Research (BMBF), Grant No. 03SF0451, in the framework of HEMCP are acknowledged. F.D. gratefully acknowledges financial support from Alexey Chernikov and the Würzburg-Dresden Cluster of Excellence on Complexity and Topology in Quantum Matter ct.qmat (EXC 2147, Project-ID 390858490). F.L. thanks the financial support from China Scholarship Council (File No. 202108440218) for his stay in Germany. Z.S. was supported by ERC-CZ program (project LL2101) from Ministry of Education Youth and Sports (MEYS) and used large infrastructure from project reg. No. CZ.02.1.01/0.0/0.0/15_003/0000444 financed by the EFRR.

# Supplemental Figures

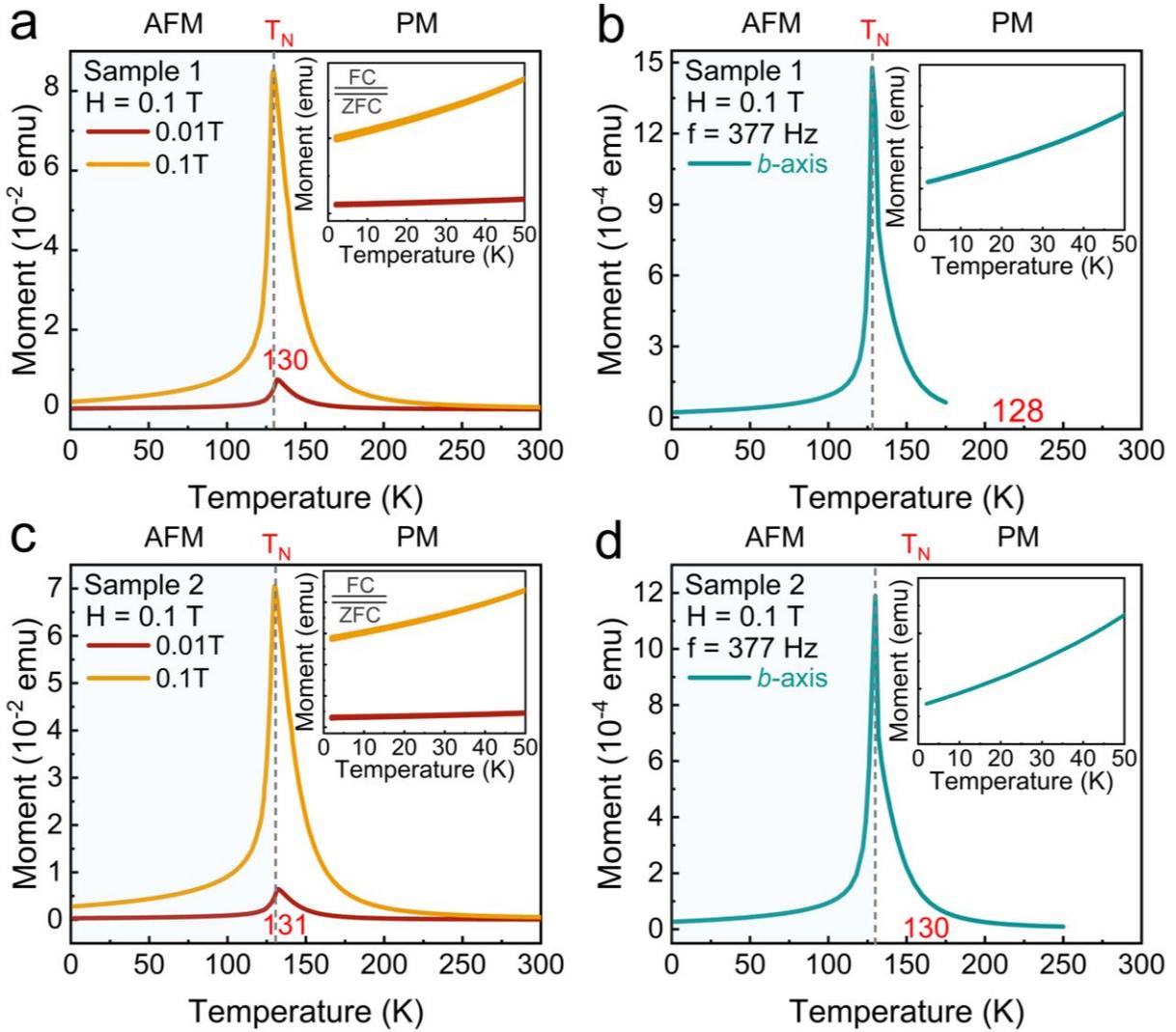

**FIG. S1. Magnetic properties of CrSBr.** a,c) Magnetic moment and b,d) AC magnetic moment of randomly selected CrSBr. The insets show the zoom-in of the low-temperature data.

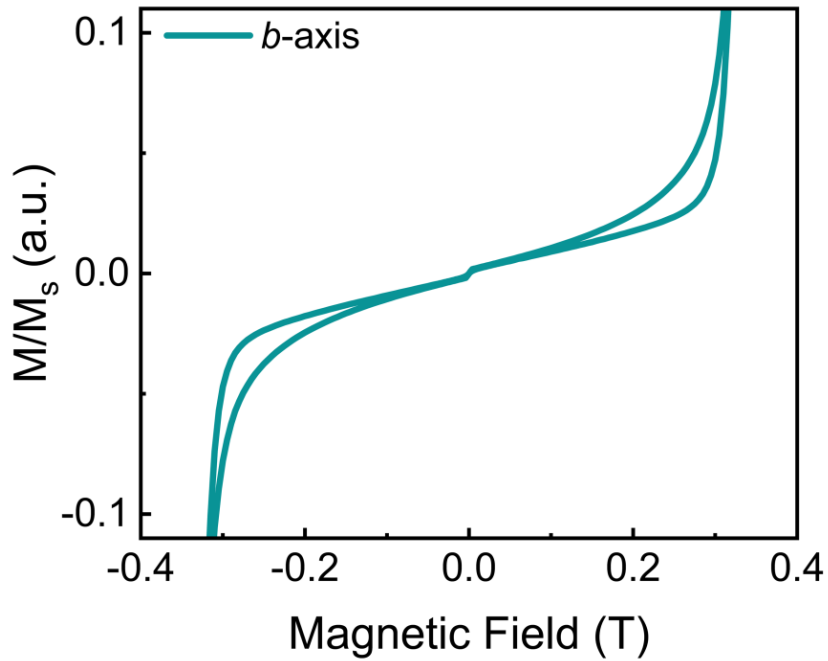

**FIG. S2.** Magnetization of CrSBr at small values of the applied magnetic field.

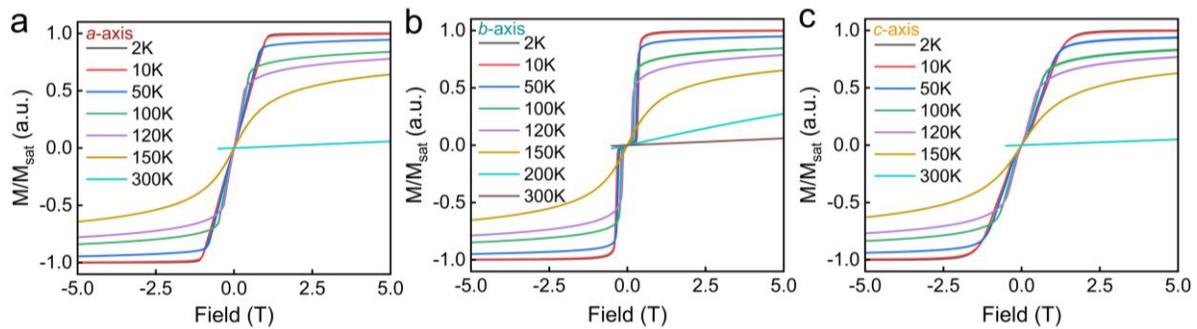

**FIG. S3. Magnetic properties of CrSBr.** Magnetization versus magnetic field along the (a) *a*-axis, (b) *b*-axis, and (c) *c*-axis at different temperatures.